\documentclass[
reprint,
superscriptaddress,
amsmath,amssymb,
aps,
prd,
floatfix,
]{revtex4-1}

\usepackage{graphicx}
\usepackage{dcolumn}
\usepackage{bm}
\usepackage{xcolor}

\usepackage{amsfonts}
\usepackage{booktabs}
\usepackage{float}
\usepackage{units}
\usepackage{hyperref}
\usepackage{lineno}

\newcommand{\gammaray}{\mbox{$\gamma$-ray}}

\begin{document}

\title{Combined search for neutrinos from dark matter self-annihilation in the Galactic Centre with ANTARES and IceCube}

%--------------------------
%ANTARES
%--------------------------
\author{A.~Albert}
\affiliation{Universit\'e de Strasbourg, CNRS,  IPHC UMR 7178, F-67000 Strasbourg, France}
\affiliation{Universit\'e de Haute Alsace, F-68200 Mulhouse, France}
\author{M.~Andr\'e}
\affiliation{Technical University of Catalonia, Laboratory of Applied Bioacoustics, Rambla Exposici\'o, 08800 Vilanova i la Geltr\'u, Barcelona, Spain}
\author{M.~Anghinolfi}
\affiliation{INFN - Sezione di Genova, Via Dodecaneso 33, 16146 Genova, Italy}
%\author{G.~Anton}
%\affiliation{Friedrich-Alexander-Universit\"at Erlangen-N\"urnberg, Erlangen Centre for Astroparticle Physics, Erwin-Rommel-Str. 1, 91058 Erlangen, Germany}
\author{M.~Ardid}
\affiliation{Institut d'Investigaci\'o per a la Gesti\'o Integrada de les Zones Costaneres (IGIC) - Universitat Polit\`ecnica de Val\`encia. C/  Paranimf 1, 46730 Gandia, Spain}
\author{J.-J.~Aubert}
\affiliation{Aix Marseille Univ, CNRS/IN2P3, CPPM, Marseille, France}
\author{J.~Aublin}
\affiliation{APC, Universit\'e de Paris, CNRS, Astroparticule et Cosmologie, F-75013 Paris, France}
\author{B.~Baret}
\affiliation{APC, Universit\'e de Paris, CNRS, Astroparticule et Cosmologie, F-75013 Paris, France}
\author{S.~Basa}
\affiliation{Aix Marseille Univ, CNRS, CNES, LAM, Marseille, France}
\author{B.~Belhorma}
\affiliation{National Center for Energy Sciences and Nuclear Techniques, B.P.1382, R. P.10001 Rabat, Morocco}
\author{V.~Bertin}
\affiliation{Aix Marseille Univ, CNRS/IN2P3, CPPM, Marseille, France}
% ...

\collaboration{ANTARES Collaboration}
\noaffiliation

%------------------------------
%IceCube
%------------------------------

\author{M. G. Aartsen}
\affiliation{Dept. of Physics and Astronomy, University of Canterbury, Private Bag 4800, Christchurch, New Zealand}
\author{M. Ackermann}
\affiliation{DESY, D-15738 Zeuthen, Germany}
\author{J. Adams}
\affiliation{Dept. of Physics and Astronomy, University of Canterbury, Private Bag 4800, Christchurch, New Zealand}
\author{J. A. Aguilar}
\affiliation{Universit{\'e} Libre de Bruxelles, Science Faculty CP230, B-1050 Brussels, Belgium}
\author{M. Ahlers}
\affiliation{Niels Bohr Institute, University of Copenhagen, DK-2100 Copenhagen, Denmark}
\author{M. Ahrens}
\affiliation{Oskar Klein Centre and Dept. of Physics, Stockholm University, SE-10691 Stockholm, Sweden}
\author{C. Alispach}
\affiliation{D{\'e}partement de physique nucl{\'e}aire et corpusculaire, Universit{\'e} de Gen{\`e}ve, CH-1211 Gen{\`e}ve, Switzerland}
\author{K. Andeen}
\affiliation{Department of Physics, Marquette University, Milwaukee, WI, 53201, USA}
\author{T. Anderson}
\affiliation{Dept. of Physics, Pennsylvania State University, University Park, PA 16802, USA}
\author{I. Ansseau}
\affiliation{Universit{\'e} Libre de Bruxelles, Science Faculty CP230, B-1050 Brussels, Belgium}
% ...

\collaboration{IceCube Collaboration}
\noaffiliation

\date{\today}

\begin{abstract}
We present the results of the first combined dark matter search targeting the Galactic Centre using the ANTARES and IceCube neutrino telescopes. For dark matter particles with masses from 50 to 1000\,GeV, the sensitivities on the self-annihilation cross section set by ANTARES and IceCube are comparable, making this mass range particularly interesting for a joint analysis. Dark matter self-annihilation through the $\tau^+\tau^-$, $\mu^+\mu^-$, $b\bar{b}$ and $W^+W^-$ channels is considered for both the Navarro-Frenk-White and Burkert halo profiles. In the combination of 2,101.6 days of ANTARES data and 1,007 days of IceCube data, no excess over the expected background is observed. Limits on the thermally-averaged dark matter annihilation cross section $\langle\sigma_A\upsilon\rangle$ are set. These limits present an improvement of up to a factor of two in the studied dark matter mass range with respect to the individual limits published by both collaborations. When considering dark matter particles with a mass of 200\,GeV annihilating through the $\tau^+\tau^-$ channel, the value obtained for the limit is $7.44 \times 10^{-24} \text{cm}^{3}\text{s}^{-1}$ for the Navarro-Frenk-White halo profile. For the purpose of this joint analysis, the model parameters and the likelihood are unified, providing a benchmark for forthcoming dark matter searches performed by neutrino telescopes.
\end{abstract}

\maketitle

\section{Introduction}\label{Intro}
Dark matter was first postulated in the 1930s and its existence has been established by a wealth of astrophysical as well as cosmological observations, both at Galactic and extragalactic scales~\cite{EarlyDMHistory, Planck}. Nevertheless, the nature of dark matter remains largely unknown and a variety of theoretical models are considered in order to solve this mystery~\cite{DM_evidence}. A common hypothesis assumes dark matter to be composed of, yet unobserved, Weakly Interactive Massive Particles (WIMPs)~\cite{WIMP_constraint}. Searches for dark matter are typically carried out in three different ways: direct detection of nuclear recoil from WIMP-nucleus interactions~\cite{DirectDetection}, dark matter production in particle accelerators~\cite{DMproduction} and indirect searches~\cite{HESS, Veritas, Fermi-MAGIC}. When annihilating or decaying, dark matter particles are expected to produce Standard Model particles. 
These will eventually yield stable charged particles present in the cosmic radiations, as well as neutrinos and \gammaray{s}.
Indirect searches look for these messengers, which can be detected by space or ground-based observatories.

Observations of the kinematics of stars and N-body simulations suggest that galaxies and galaxy clusters are embedded in dark matter halos, with an increased density towards the centre~\cite{Moore, Kravtsov}. In addition, dark matter particles are expected to accumulate gravitationally at the centre of massive objects, such as the Earth~\cite{IceCube_Earth, ANTARES_Earth} and the Sun~\cite{IceCube_Sun, ANTARES_Sun, ANTARES_Schecluded}, after losing energy via scattering. The enhanced concentration of dark matter at the centre of these objects would favour their annihilation into secondary particles, making massive objects good targets for indirect searches.

The analysis presented in this paper consists in a search for neutrinos from dark matter self-annihilation in the centre of the Milky Way. 
In this paper, the term ``neutrino" refers to $\nu+\bar\nu$ since the events generated by neutrinos and anti-neutrinos are seen indistinguishably in the two detectors considered.
Corresponding limits on the thermally-averaged annihilation cross section, $\langle\sigma_A\upsilon\rangle$, have already been set by the ANTARES and IceCube collaborations~\cite{ANTARES_GCWIMP, ANTARES_GCWIMP_old, IC86_GCWIMP, GalacticHalo, MultiPole}. Both neutrino telescopes are optimised for the detection of high-energy neutrinos ($\sim$1\,TeV). For dark matter masses ranging from 50 to 1000\,GeV, the limits obtained by the two telescopes are comparable, which makes this region interesting for a joint analysis. By combining the datasets of both experiments, the goal is to improve the detection potential in this particular mass range. In order to perform this combined search, an important aspect was to identify the differences in the methods used by the two collaborations and to reconcile them.

This paper is structured as follows. The expected neutrino flux from dark matter annihilation is discussed in Section~\ref{IndirectSearch}. In Section~\ref{Detectors}, the ANTARES and IceCube neutrino detectors are presented. Section~\ref{EventSelection} gives an overview of the datasets used for the combined search. The analysis method is introduced in Section~\ref{Method}. In section~\ref{Systematics}, the systematic uncertainties are addressed. Finally, the results are shown and discussed in Section~\ref{Results}.

%---------------------------------------------------------
\section{Indirect dark matter search with neutrinos}\label{IndirectSearch}
%---------------------------------------------------------

The expected differential flux of secondary neutrinos from dark matter self-annihilation in the Galactic Centre is defined following reference~\cite{NeutrinoFluxDM}:

\begin{equation}
    \frac{\mathrm{d}\phi_{\mathrm{\nu}}}{\mathrm{d}E_{\mathrm{\nu}}} = \frac{1}{4\pi} \, \frac{\langle \sigma_A \upsilon \rangle}{2 \, m_{\mathrm{DM}}^2} \; \frac{\mathrm{d}N_{\mathrm{\nu}}}{\mathrm{d}E_{\mathrm{\nu}}} \; J \, ,
    \label{eq:sig_expectation}
\end{equation}

\noindent where $\langle\sigma_A\upsilon\rangle$ is the thermally-averaged self-annihilation cross section, $m_{\mathrm{DM}}$ is the mass of the dark matter particle and $\mathrm{d}N_{\mathrm{\nu}}/\mathrm{d}E_{\mathrm{\nu}}$ is the differential number of neutrinos per annihilating dark matter pair. The factor $1/4\pi$ arises from the assumed spherical symmetry of the dark matter self-annihilation.
The J-factor is expressed as

\begin{equation}
    J = \int_{\Delta \Omega} \mathrm{d}\Omega(\Psi) \int_{\mathrm{l.o.s}} \rho_{\mathrm{DM}}^2\left(r(l, \Psi)\right) \, \mathrm{d}l \, , 
    \label{eq:J_factor}
\end{equation}

\noindent and is defined as the integral over the solid angle, $\Delta \Omega$, of the squared dark matter density evaluated along the line of sight (l.o.s.). The J-factor depends on the opening angle to the Galactic Centre, $\Psi$. The squared dark matter mass and dark matter density, as well as the factor $1/2$, result from the fact that two dark matter particles are needed for each annihilation.

The density distribution of dark matter in galaxies as a function of the distance $r$ to the Galactic Centre can be parameterized by an extension of the Zhao profile~\cite{DM_density}:

\begin{align}
    &\rho_{\mathrm{DM}}(r) = \frac{\rho_0}{\left(\delta + \frac{r}{r_{\mathrm{s}}}\right)^{\gamma} \cdot \left[1+\left(\frac{r}{r_{\mathrm{s}}}\right)^{\alpha}\right]^{(\beta-\gamma)/\alpha}} \; .
    \label{eq:DM_density}
\end{align}

\begin{figure*}
    \centering
    \begin{minipage}{.49\textwidth}
    \includegraphics[width=\linewidth]{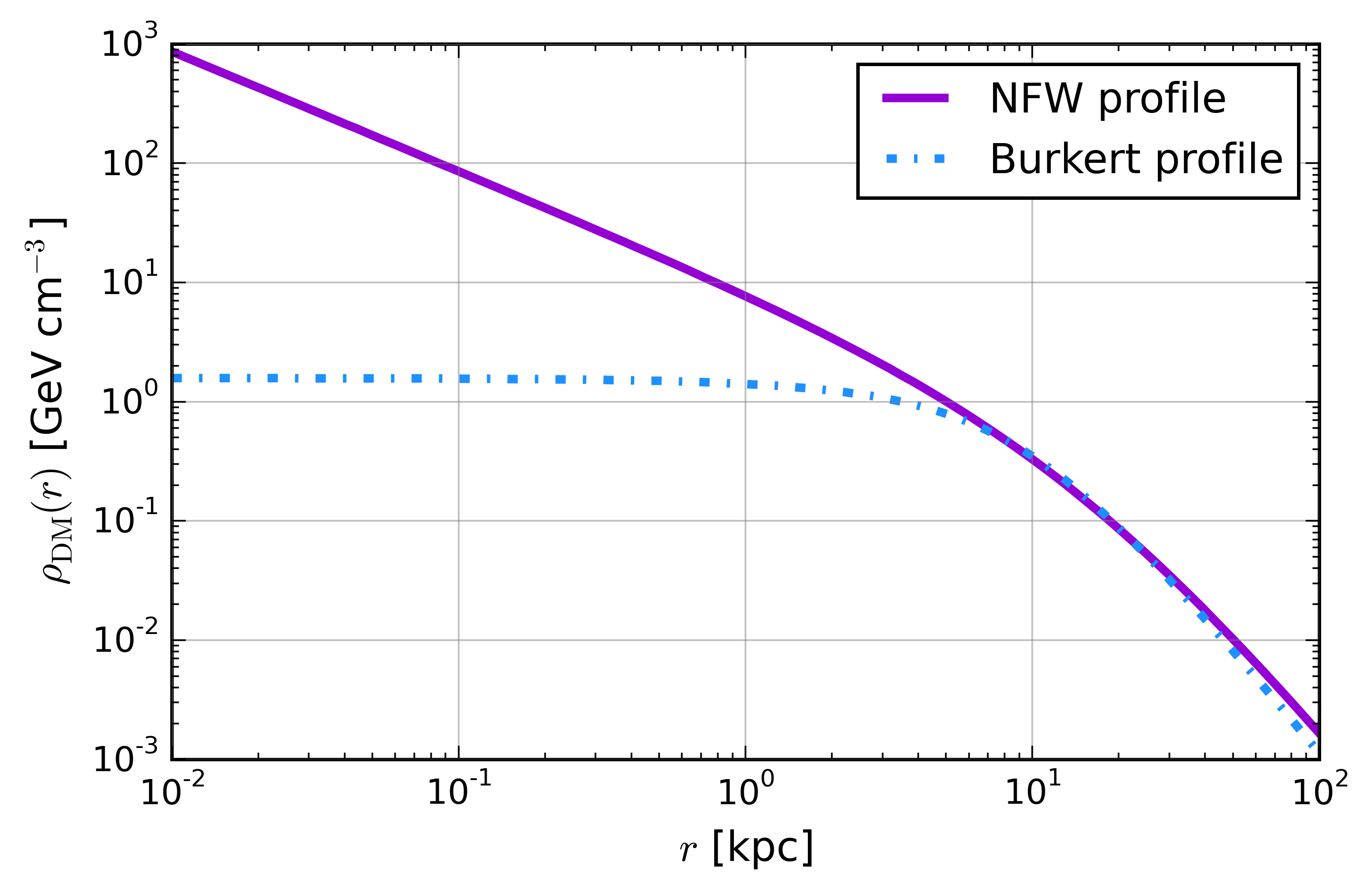}
    \end{minipage}
    \hfill
    \begin{minipage}{.49\textwidth}
    \includegraphics[width=\linewidth]{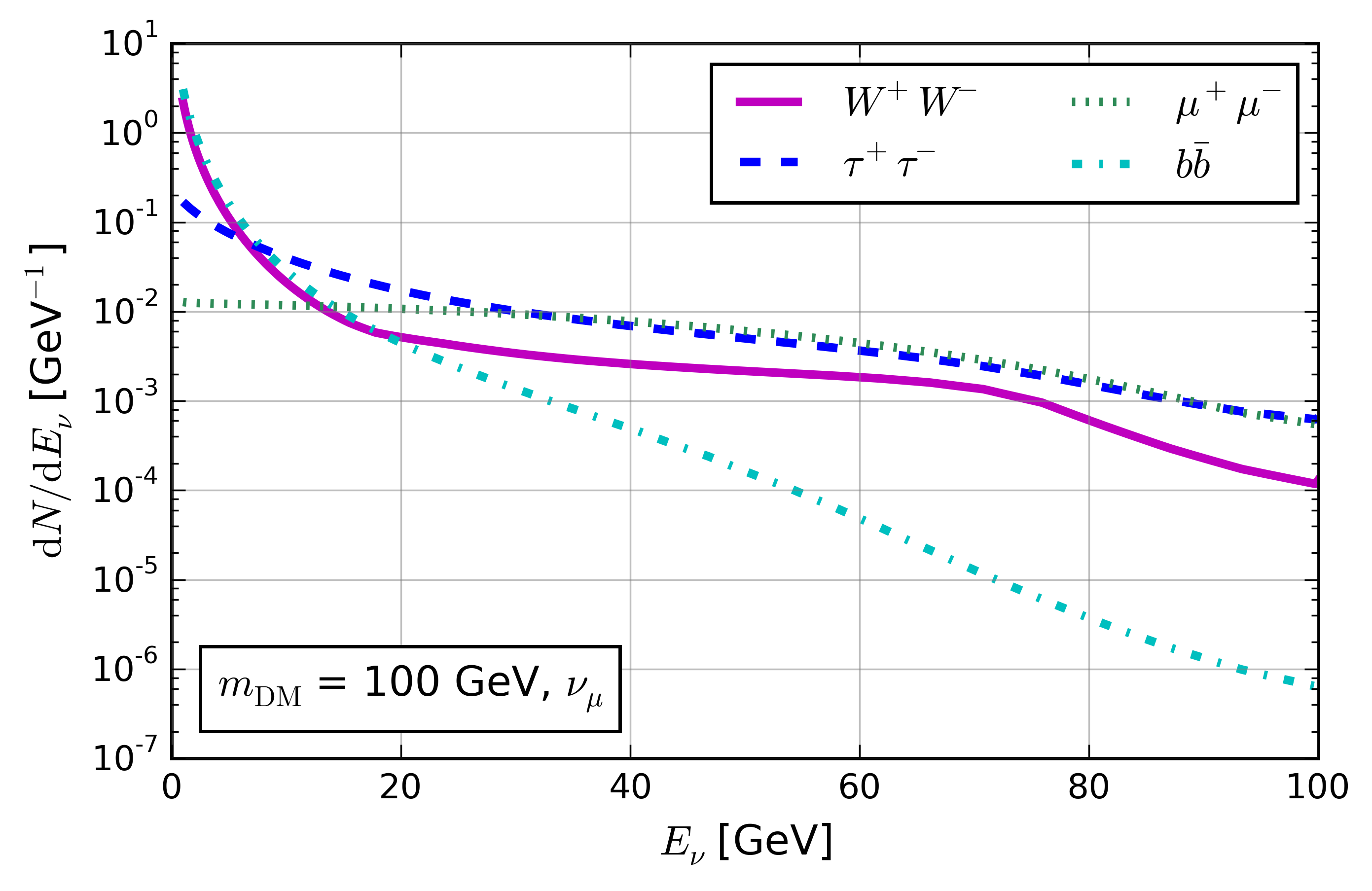}
    \end{minipage}
    \caption{\textbf{Left:} Dark matter density, $\rho_{\mathrm{DM}}(r)$, as a function of the radial distance to the Galactic Centre, r, for the NFW and Burkert profiles. \textbf{Right:} Muon neutrino spectra at Earth for a dark matter mass of 100\,GeV and the four self-annihilation channels.}
    \label{fig:Model_parameters}
\end{figure*}

\noindent Both the normalisation density, $\rho_0$, and the scale radius, $r_s$, have to be evaluated for each galaxy.
Both the ANTARES and IceCube analyses took values for these free model parameters from reference~\cite{Nesti}. For consistency reasons, these values are also used for the combined search (see Table~\ref{tab:ModelParam}).
Since the J-factor depends on the dark matter density used, we consider two dark matter halo models to account for this uncertainty. Both of them are described by Equation~\ref{eq:DM_density}, where the dimensionless parameters $(\alpha, \beta, \gamma, \delta)$ take the values (1,3,1,0) for the  Navarro-Frenk-White profile (NFW)~\cite{NFW} and (2,3,1,1) for the Burkert profile~\cite{Burkert}. 
While the two models differ by orders of magnitude close to the Galactic Centre, they become rather similar outside the solar circle, $R_{sc}$=8.5\,kpc, in agreement with uncertainty estimations from galactic rotation curves~\cite{RotationCurves}.
The resulting dark matter densities as a function of $r$ are shown in the left panel of Figure~\ref{fig:Model_parameters} for both halo profiles.

\begin{table}
\begin{center}
\begin{tabular}{|c|c|c|c|}
\hline
Parameters & Units & NFW & Burkert \\
\hline
$\rho_0$ & $10^{7}M_{\mathrm{\odot}}/\text{kpc}^3$ & $1.4$ & $4.1$  \\
\hline
$r_s$ & kpc & $16.1$ & $9.3$\\
\hline
\end{tabular}
\caption{Parameters of the dark matter halo profiles for the Milky Way taken from reference~\cite{Nesti}.}
\label{tab:ModelParam}
\end{center}
\end{table}

Along with the spatial distribution of dark matter, given by the J-factor, the spectra of secondary particles from dark matter annihilation is also a necessary theoretical input for this analysis. In our effort to combine the methods of both experiments, we found differences in the energy spectra used for previous analyses. While the spectra known as PPPC4~\cite{Cirelli} tables were used by ANTARES, IceCube used spectra computed directly with PYTHIA~\cite{Pythia}. For the purpose of the combined analysis, it was imperative to use the same spectra for both detectors. The PPPC4 tables are preferred as they take electroweak corrections into account. As a result, we noticed variations of up to 25$\%$ of the IceCube-only limits computed with the PPPC4 spectra when compared to the limits obtained with the previously used PYTHIA spectra. We consider dark matter annihilating through four self-annihilation channels. A 100$\%$ branching ratio to $W^+W^-$, $\tau^+\tau^-$, $\mu^+\mu^-$ or $b\bar{b}$ is assumed. The corresponding muon neutrino spectra at Earth for every annihilation process, $\mathrm{d}N_{\mathrm{\nu}}/\mathrm{d}E_{\mathrm{\nu}}$, are shown in the right panel of Figure~\ref{fig:Model_parameters} for a dark matter mass of 100\,GeV.

This analysis is sensitive to any dark matter candidate self-annihilating to Standard Model particles and leading to the production of neutrinos through the four channels studied. Throughout this work, dark matter masses ranging from 50 to 1000\,GeV are considered.

%---------------------------------------------------------
\section{Detectors}\label{Detectors}
%---------------------------------------------------------

Given the small interaction cross section of neutrinos, a large volume of target material is required for the neutrino detection. For Cherenkov detectors, such as ANTARES and IceCube, this was achieved by installing photomultiplier tubes (PMTs) in a transparent natural medium. These photo-sensors then record the Cherenkov emission induced by secondary charged particles produced by the interaction of neutrinos in the surrounding environment.

ANTARES is an underwater neutrino telescope deployed in the Mediterranean Sea, 40\,km offshore from Toulon (France) at coordinates 42$^{\circ}$48'N, 6$^{\circ}$10'E~\cite{ANTARES}. The detector is composed of 12 vertical detection lines, horizontally spaced by 70 metres. Each string holds 25 storeys of 3 photo-detectors separated vertically by 14.5 metres. The strings are anchored to the seabed at a depth of 2,475 metres, covering a volume of more than 0.01\,km$^3$.

IceCube is a cubic-kilometre neutrino telescope located at the geographic South Pole~\cite{IceCube}. The detector consists of 5,160 PMTs attached to vertical strings disseminated in 86 boreholes~\cite{IceCube_DOM}, between depths of 1,450 to 2,500 metres. The IceCube array is composed of 86 strings instrumented with 60 Digital Optical Modules (DOMs). Among them, 78 strings are arranged on a hexagonal grid with a spacing of 125 metres, with a vertical separation of 17 metres between each DOM. The eight remaining strings are deployed more compactly at the centre of the array, forming the DeepCore sub-detector~\cite{DeepCore}. A horizontal distance of 72 metres separates the DeepCore strings with a vertical spacing of 7 metres between each DOM. The fiducial volume of DeepCore forms a 125 metres radius by 350 metres long cylinder, which includes seven regular IceCube strings.

%---------------------------------------------------------
\section{Event selection}\label{EventSelection}
%---------------------------------------------------------

This joint analysis makes use of individual datasets which were designed for previous analyses of the corresponding collaborations. Both samples are optimised to search for dark matter in the Galactic Centre. Considering the different scale and location of the two detectors, distinct methods were used to reduce the background. The main backgrounds of neutrino telescopes consist of atmospheric muons and neutrinos produced by the interaction of cosmic rays with nuclei in the upper atmosphere. The contribution from atmospheric muons is six orders of magnitude larger than the background from atmospheric neutrinos. However, in the up-going direction, muons are suppressed as they are filtered out by the Earth.

The Galactic Centre is located in the Southern Hemisphere, at declination $\delta_{\mathrm{GC}} \sim -29.01^{\circ}$. Since declinations between $0^{\circ}$ and $-90^{\circ}$ are always above the horizon of IceCube, events coming from the Galactic Centre are seen as down-going events in the detector.
Therefore, we consider a smaller fiducial volume for this analysis, since the outer part of the detector is used as a veto to reject atmospheric muons. The effective volume is reduced to the 8 DeepCore strings and the 7 surrounding IceCube strings. In addition, only DOMs with depths between 2140 and 2420\,m are considered. 
Hence, the resulting effective volume for IceCube is about $0.015 \, \mathrm{km}^{3}$, which is comparable to the ANTARES instrumented volume.
Unlike IceCube, ANTARES does not have a fixed view of the Galactic Centre in local coordinates. Hence, declinations below $-47^{\circ}$ are favoured since they are always seen as up-going in the ANTARES detector, while events with declinations between $-47^{\circ}$ and $47^{\circ}$ are below the horizon for only a part of the sidereal day. As a result, ANTARES has a visibility of the Galactic Centre at about $75\%$ of the time and no instrumental veto is required for this analysis.

The ANTARES dataset consists of events recorded over 9 years between 2007 and 2015, resulting in an effective livetime of 2,101.6 days. This sample is composed of up-going track-like events and was optimised for a previous dark matter search based on the same dataset~\cite{ANTARES_GCWIMP}. According to the number of strings with triggered PMTs, two different reconstruction algorithms are used. The single-line reconstruction (QFit)~\cite{ANTARES_QFit}, which is optimised for energies below 100\,GeV, can reconstruct only the zenith angle of the events. At higher neutrino energies, the multi-line algorithm ($\lambda$Fit) is used~\cite{ANTARES_LambdaFit} since PMTs from more than one string are likely to be triggered. Both algorithms are characterised by a parameter representing the quality of the reconstructed track. The final selection results in 1,077 reconstructed neutrino events for QFit and 15,651 events for $\lambda$Fit. Since these cuts strongly favour the reconstruction of muon tracks produced in the charged-current interaction of muon neutrinos, only neutrinos of this flavour are considered.

For IceCube, a data sample thoroughly described in reference~\cite{IC86_GCWIMP} is used. That selection consists of events recorded from 2012 to 2015 with the 86-string configuration, for a total livetime of 1,007 days.
The purpose was to select track-like events starting within the detector volume. Such events originate mainly from the charged-current interactions of muon neutrinos within the detector. Unlike ANTARES, this event selection does not completely reject non-muon neutrinos. Therefore, electron or tau neutrinos with similar topology remains in the sample. The final selection results in a total of 22,622 events.

%---------------------------------------------------------
\section{Analysis procedure}\label{Method}
%---------------------------------------------------------

A binned likelihood method is applied in order to search for an excess of signal neutrinos from the Galactic Centre. In this approach, the distribution of the data is compared to what is expected from the background and signal distributions for given combinations of halo profile, dark matter mass and annihilation channel. The information about the shape of the signal and background is contained in probability density functions (PDFs). Likelihood functions are defined for each experiment, with PDFs built differently for ANTARES and IceCube.

\begin{figure}
    \centering
    \begin{minipage}{\linewidth}
    \includegraphics[width=\linewidth]{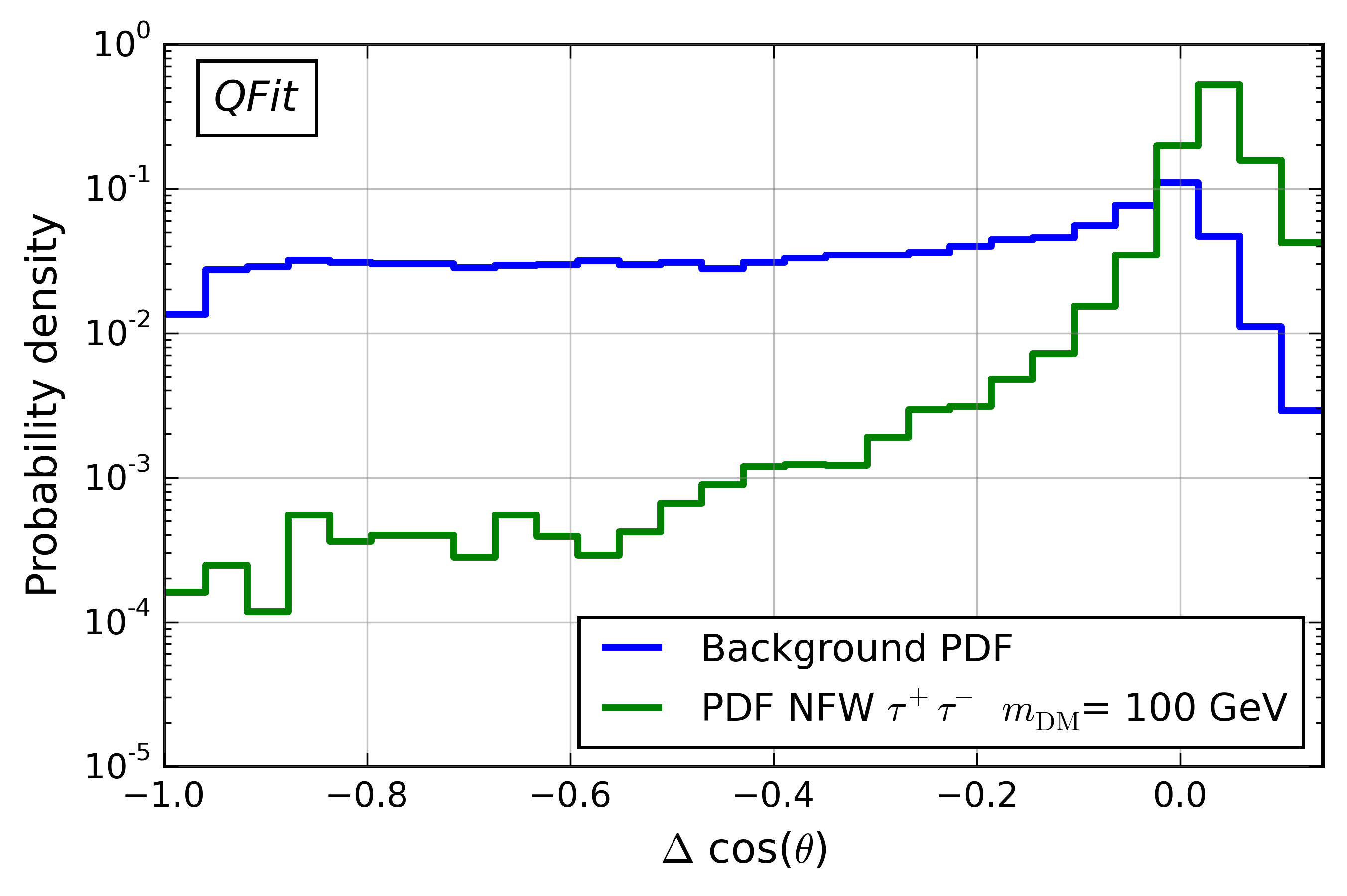}
    \end{minipage}
    \hfill
    \begin{minipage}{\linewidth}
    \includegraphics[width=\linewidth]{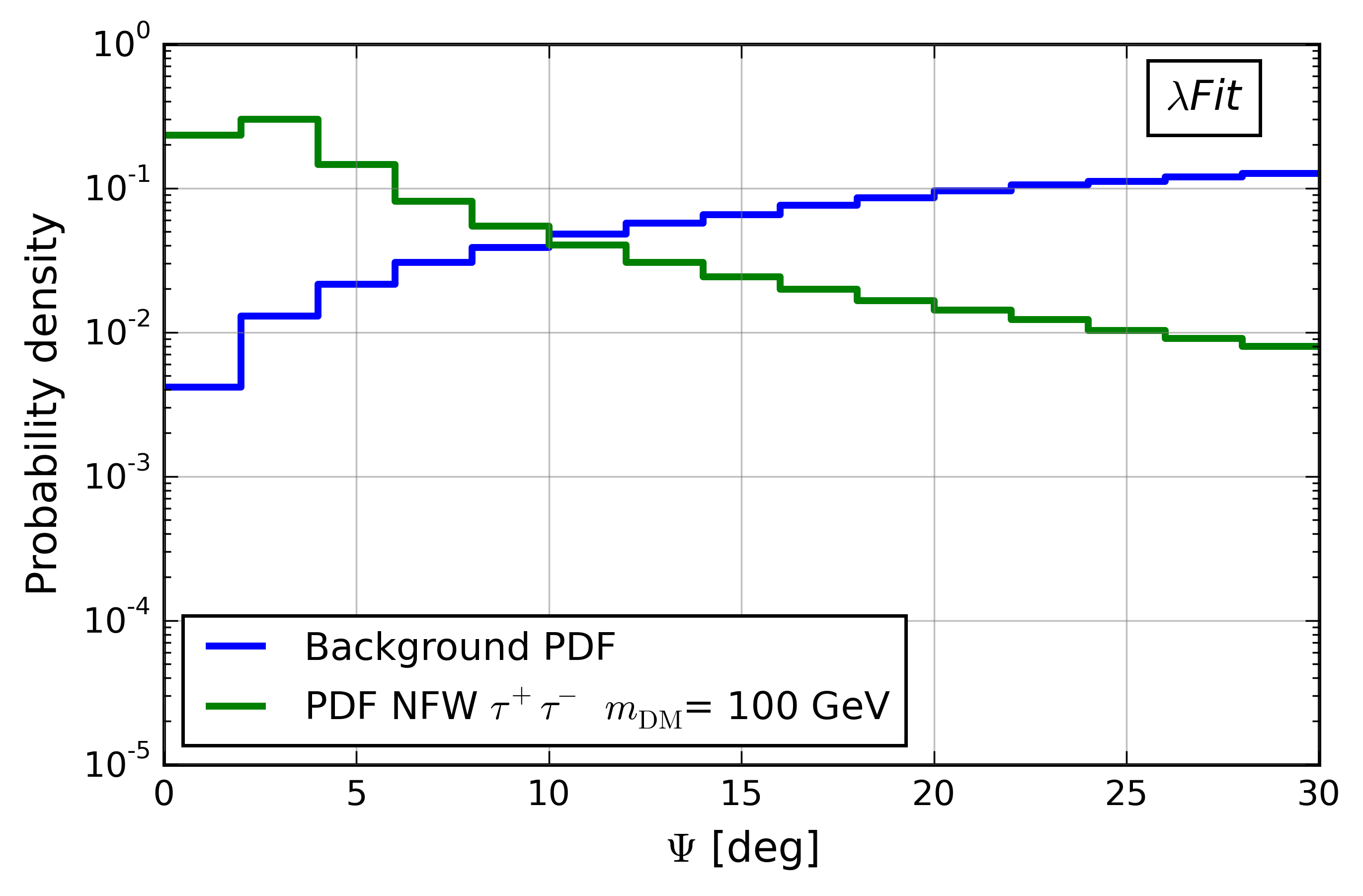}
    \end{minipage}
    \caption{\textbf{Top:} ANTARES PDFs for the QFit reconstruction. \textbf{Bottom:} ANTARES PDFs for the $\lambda$Fit reconstruction. Both histograms show the background (blue) and signal (green) PDFs for the $\tau^+\tau^-$ annihilation channel and NFW profile, assuming $m_{\mathrm{DM}}$ = 100\,GeV.}
    \label{fig:ANTARES_PDFs}
\end{figure}

\begin{figure*}
    \centering
    \begin{minipage}{0.49\linewidth}
    \includegraphics[width=\linewidth]{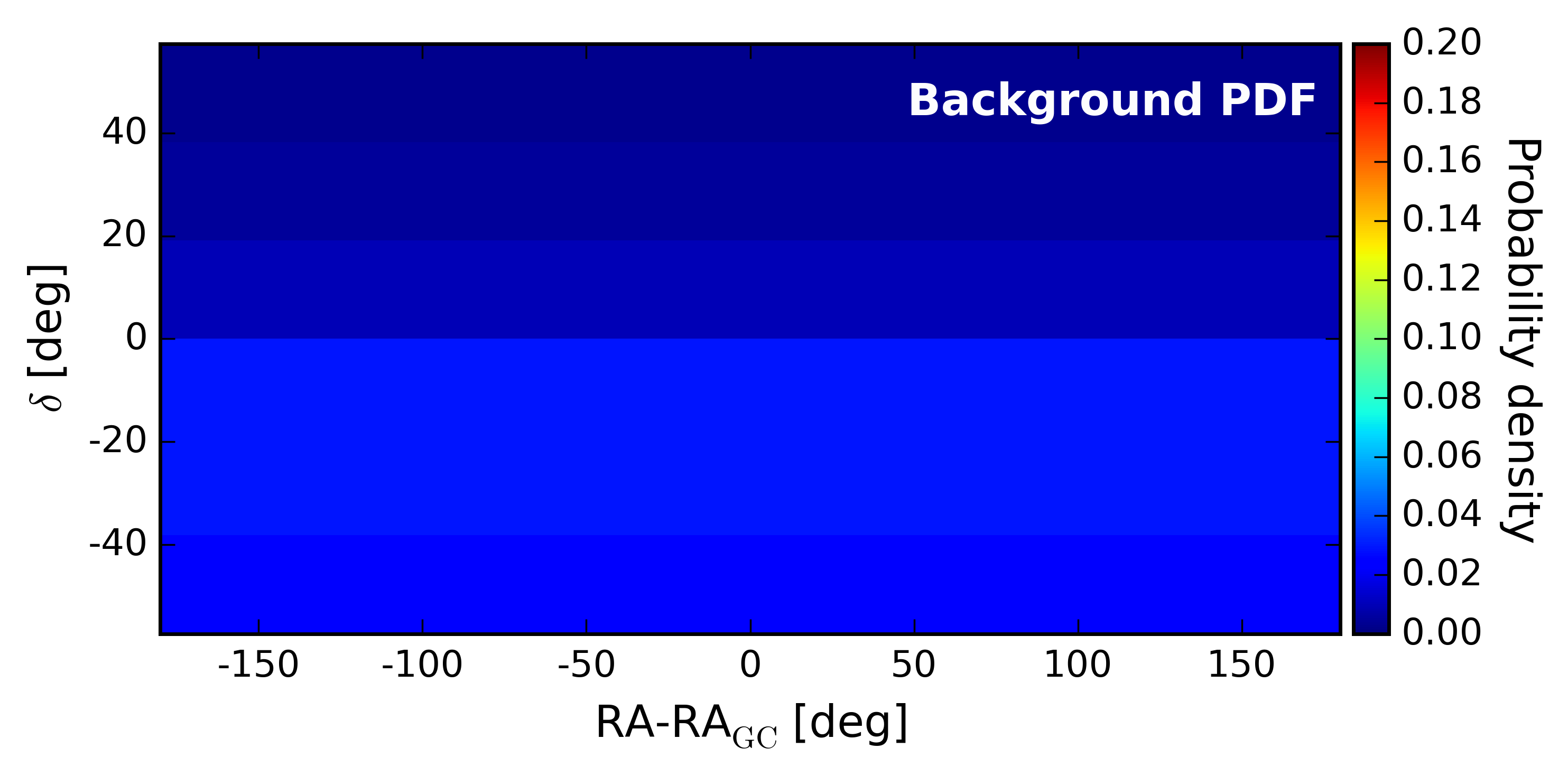}
    \end{minipage}
    \hfill
    \begin{minipage}{0.49\linewidth}
    \includegraphics[width=\linewidth]{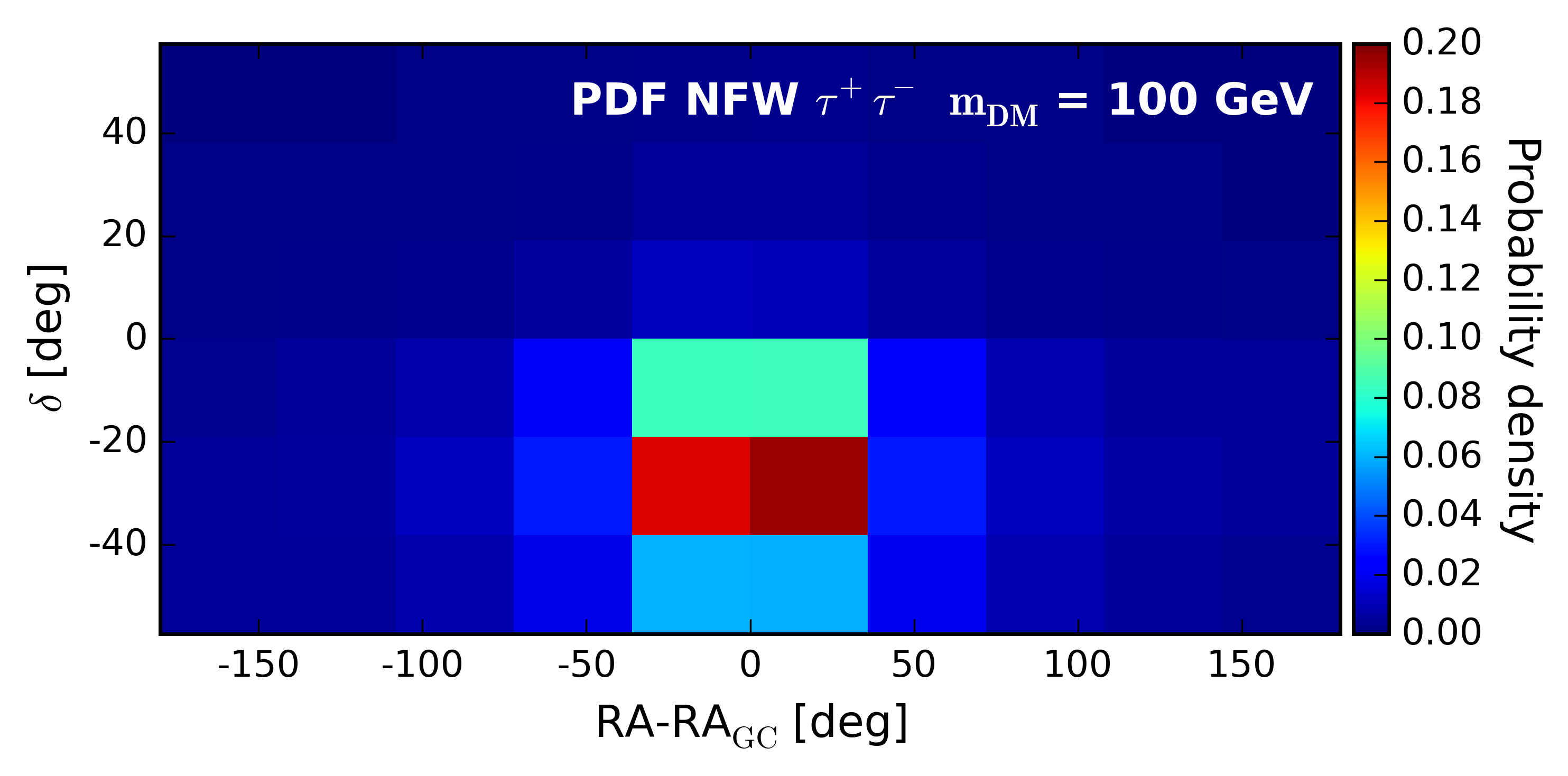}
    \end{minipage}
    \caption{\textbf{Left:} IceCube background PDF obtained from data scrambled in RA, where the colour scale expresses the probability density. \textbf{Right:} IceCube signal PDF for $\tau^+\tau^-$ channel and $m_{\mathrm{DM}}$ = 100\,GeV assuming the NFW profile.
    }
    \label{fig:IceCube_PDFs}
\end{figure*}

The ANTARES PDFs represent the angular distance of each event from the source. For QFit, we use 28 bins in $\Delta \cos(\theta) = \cos(\theta_{\mathrm{GC}}) - \cos(\theta_{\mathrm{event}})$ from -1 to 0.14, where $\cos(\theta_{\mathrm{event}})$ is the zenith of the reconstructed event track and $\cos(\theta_{\mathrm{GC}})$ represents the zenith position of the Galactic Centre at the time of the event (see top panel of Figure~\ref{fig:ANTARES_PDFs}). For $\lambda$Fit, we consider 15 bins in $\Psi$ ranging from 0$^{\circ}$ to 30$^{\circ}$, where $\Psi$ is the space angle between the Galactic Centre and the event track (see bottom panel of Figure~\ref{fig:ANTARES_PDFs}). In the case of IceCube, 2-dimensional distributions are used (see Figure~\ref{fig:IceCube_PDFs}). The binning consists of 6 bins in declination ranging from -1 to 1\,rad and 10 bins in right ascension (RA) covering the range from -$\pi$ to $\pi$\,rad.

 For both experiments, the signal PDFs are estimated from generic samples of simulated neutrinos, which are then weighted with the source morphology and the neutrino spectrum for each halo profile, dark matter mass and annihilation channel. Assuming uniformity of the background in RA, the IceCube background PDF is determined by scrambling the data in RA and subtracting the expected signal. For ANTARES, the $\lambda$Fit background PDF is also determined from experimental data scrambled in RA, while the QFit background PDF is obtained by scrambling the arrival time of the events.

Any given event distribution, $f^{i}(\mu)$, can be expressed as a superposition of the signal, $f^{i}_{\mathrm{s}}$, and background, $f^{i}_{\mathrm{bg}}$, PDFs:

\begin{equation}
    f^{i}(\mu) = \mu \, f_s^i \, + \, (1 - \mu) \, f^i_{\mathrm{bg}} \, ,
    \label{eq:Signal_fraction}
\end{equation}

\noindent where $\mu \in [0,1]$ is the fraction of signal events assumed to be present in the total sample. 

The likelihood is defined as the product of the Poisson probabilities to observe $n_{\mathrm{obs}}^i$ events in a particular bin $i$:

\begin{equation}
    \mathcal{L}(\mu) = \prod_{i=min}^{max} \frac{\left(n_{\mathrm{obs}}^{\mathrm{tot}} \, f^{i}(\mu)\right)^{n^{i}_{\mathrm{obs}}}}{n^{i}_{\mathrm{obs}}!} e^{-n_{\mathrm{obs}}^{\mathrm{tot}} \, f^{i}(\mu)}\, .
    \label{eq:Likelihood}
\end{equation}

\noindent The number of observed events in a bin $i$, $n^{i}_{\mathrm{obs}}$, is compared to the expected number of events in that particular bin, given the total number of event in the data, $n_{\mathrm{obs}}^{\mathrm{tot}}$ times the fraction of events within a specific bin, $f^{i}(\mu)$, for a given value of $\mu$.

\begin{center}
\begin{figure}
\includegraphics[width=0.99\linewidth]{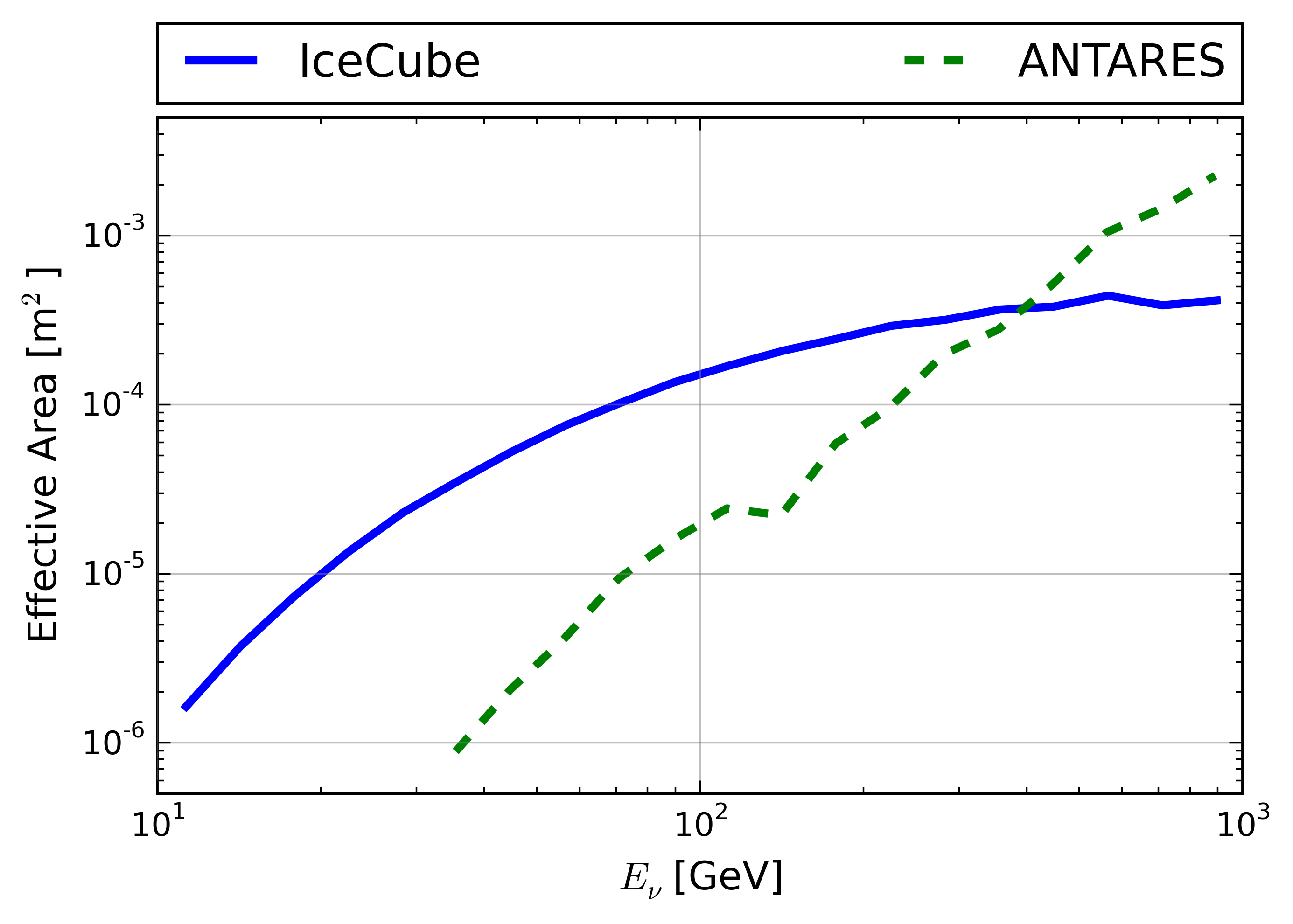}
\caption{Comparison of the effective area of the ANTARES and IceCube samples as a function of the neutrino energy for events with declination $\delta \in [\delta_{\mathrm{GC}}-30^{\circ}, \delta_{\mathrm{GC}}+ 30^{\circ}]$.}
\label{fig:EffectiveArea}
\end{figure}
\end{center}

Once defined for ANTARES and IceCube separately, the likelihoods are merged into a single combined likelihood defined as

\begin{equation}
    \mathcal{L}_{\mathrm{comb}}(\mu) = \prod_{k=A,I} \, \mathcal{L}_{k}(w_{k} \, . \, \mu) \, ,
    \label{eq:Combined_likelihood}
\end{equation}

\noindent where the index \textit{k} = (\textit{A}, \textit{I}) refers to ANTARES and IceCube, respectively. Since the signal acceptances, $\eta_{\mathrm{sig}}^k$, for a given dark matter signal (mass, annihilation channel and halo profile) are different for the two experiments, the signal fraction is weighted with a relative weight, $w_{k}$. This weight represents the relative signal acceptance of each experiment with respect to the contribution from the total event sample:

\begin{equation}
    w_{k} = \, \frac{\eta_{\mathrm{sig}}^k/\eta_{\mathrm{sig}}}{N_{\mathrm{tot}}^{k}/N_{\mathrm{tot}}} \, ,
\end{equation}

\noindent where $N_{\mathrm{tot}}$ denotes the total number of background events and is obtained by summing $N_{\mathrm{tot}}^A$ and $N_{\mathrm{tot}}^I$. The total signal acceptance, $\eta_{\mathrm{sig}}$, is defined as the sum of the individual signal acceptances, $\eta_{\mathrm{sig}}^k$, which we define as

\begin{equation}
    \eta_{\mathrm{sig}}^k = \frac{1}{8\pi} \, \frac{J}{\, T_{\mathrm{live}}^{k}}{m^{2}_{\mathrm{DM}}} \, \int{A^{k}_{\mathrm{eff}}(E) \, \frac{\mathrm{d}N_{\mathrm{\nu}}}{\mathrm{d}E_{\mathrm{\nu}}} \, \mathrm{d}E} \, ,
\end{equation}

\noindent where $T^{k}_{\mathrm{live}}$ is the experiment livetime and $A^{k}_{\mathrm{eff}}$ is the effective area of the detector. The signal acceptances are computed for each combination of dark matter mass, annihilation channel and halo profile. The effective area, computed using Monte Carlo simulations, depends on several factors such as the neutrino cross-section, the range of secondary particles, the detector efficiencies, and the selection criteria for each sample. A comparison of the effective area of the ANTARES and IceCube samples is shown in Figure~\ref{fig:EffectiveArea} for declinations between $\delta_{\mathrm{GC}}-30^{\circ}$ and $\delta_{\mathrm{GC}}+ 30^{\circ}$.
The different behaviour of the two curves can be explained by the fact that, while IceCube is limited to its vetoed fiducial volume in order to limit the background from atmospheric muons, ANTARES does not need to restrict itself to events starting within the detector volume. Therefore, the IceCube effective area hits a plateau at higher energies while ANTARES can extend its fiducial volume beyond the detector boundaries. For ANTARES, the transition between the QFit and the $\lambda$Fit reconstructions is visible around 130 GeV.

 With this likelihood method, we can obtain the best estimate of the signal fraction, $\mu_{\mathrm{best}}$, which is the value of $\mu$ maximising the likelihood, $\mathcal{L}(\mu)$. In order to evaluate the sensitivity of this analysis, we generate 100,000 pseudo-experiments sampled from the background-only PDF. For each of these pseudo-experiments, we compute the upper limit at the $90\%$ confidence level (CL), $\mu_{\mathrm{90}}$, according to the unified approach of Feldman $\&$ Cousins~\cite{Feldman-cousins}. The final sensitivity, $\hat{\mu}_{\mathrm{90}}$, is defined as the median value of these upper limits. The $\mu_{\mathrm{90}}$ distribution of the pseudo-experiments is also used to determine the statistical uncertainty of the sensitivity, which we express in terms of 1$\sigma$ and 2$\sigma$ uncertainties. 
 
 The same method is used to determine $\mu_{\mathrm{best}}$ for the unblinded data. If the obtained value is consistent with the background-only hypothesis, the corresponding upper limit on the signal fraction, $\mu_{90}$, is computed. We can then deduce the limit on the dark matter self-annihilation cross section using the relation
 
 \begin{equation}
    \langle \sigma_\mathrm{A} \upsilon \rangle =  \frac{\mu_{\mathrm{90}} \, N_{\mathrm{tot}}}{\eta_{\mathrm{sig}}} \, ,
    \label{eq:sigma_weight}
\end{equation}

\noindent for a given dark matter mass, annihilation channel and halo profile.
 
%---------------------------------------------------------
\section{Systematic uncertainties}\label{Systematics}
%---------------------------------------------------------

The sources of uncertainties can be split into theoretical and detector-related systematic uncertainties. Since the background PDFs are obtained from data for both experiments, systematic effects were only studied for the signal simulation.

For ANTARES, the uncertainty on the track direction is the dominant systematic uncertainty. To account for this, the approach used in previous ANTARES point source searches is applied. 
The determination of track parameters relies on the time resolution of the detector units, affected by the PMT transit time spread, errors in the calibration of the timing system and possible spatial misalignment of the detector lines.  As reported in reference~\cite{ANTARES_LambdaFit}, these uncertainties overall affect the angular resolution for tracks by about 15$\%$. This uncertainty is implemented in the analysis by smearing the signal PDFs by 15$\%$.

Similarly, the dominant source of systematic uncertainty of the IceCube detector results from the uncertainty on the angular resolution, which is affected by the modelling of the ice properties and the photon detection efficiency of the DOMs. These effects were studied using Monte Carlo simulations for which a variations of $\pm 1\sigma$ on the baseline set values were applied. This results in a 5-15$\%$ uncertainty from the optical properties of the ice, where the scattering and absorption lengths are modified. The optical properties of the hole ice are different than the bulk ice. Due to the presence of impurities, the scattering length of the ice in the drilling holes is shorter. The treatment of the uncertainty on the scattering length results in a worsening of 25-30$\%$ of the sensitivity when increasing the scattering length considered for the hole ice. Reciprocally, the sensitivity improves by $5-10\%$ when considering a shortening of the scattering length. The uncertainty on the photon detection efficiency of the DOMs affects the sensitivity by improving or worsening it by 5 to 40$\%$. We add in quadrature the different systematic contributions to obtain the total uncertainty, assuming all systematic uncertainties to be independent. These systematic uncertainties are included in the final results by conservatively reducing the IceCube signal acceptance $\eta_{\mathrm{sig}}^{\mathrm{I}}$ by 38$\%$.

However, astrophysical uncertainties on the dark matter halo model parameters prevail over the systematic uncertainties mentioned above.
For instance, alternative estimates of these model parameters for the NFW halo profile~\cite{McMillan}, can affect the limit by up to a factor of 1.5.
To account for uncertainties linked to the dark matter halo models, we present limits for both the NFW and Burkert profiles. The impact of the halo model choice can be seen in Figure~\ref{fig:AllChannels_Limits}, where the limits for the NFW and Burkert profiles are presented.

%---------------------------------------------------------
\section{Results and discussion}\label{Results}
%---------------------------------------------------------

\begin{figure}
    \centering
    \begin{minipage}{\linewidth}
    \includegraphics[width=\linewidth]{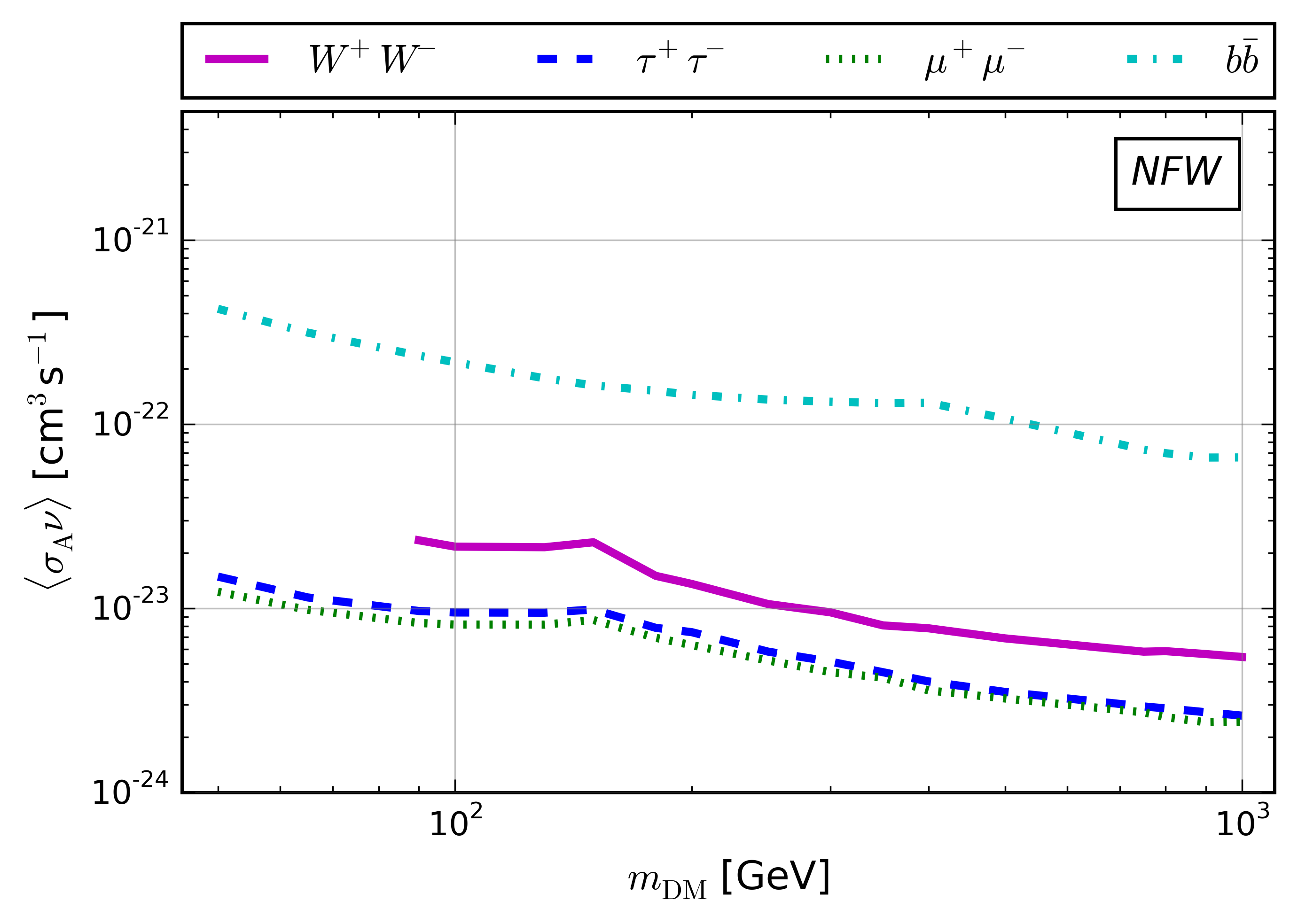}
    \end{minipage}
    \hfill
    \begin{minipage}{\linewidth}
    \includegraphics[width=\linewidth]{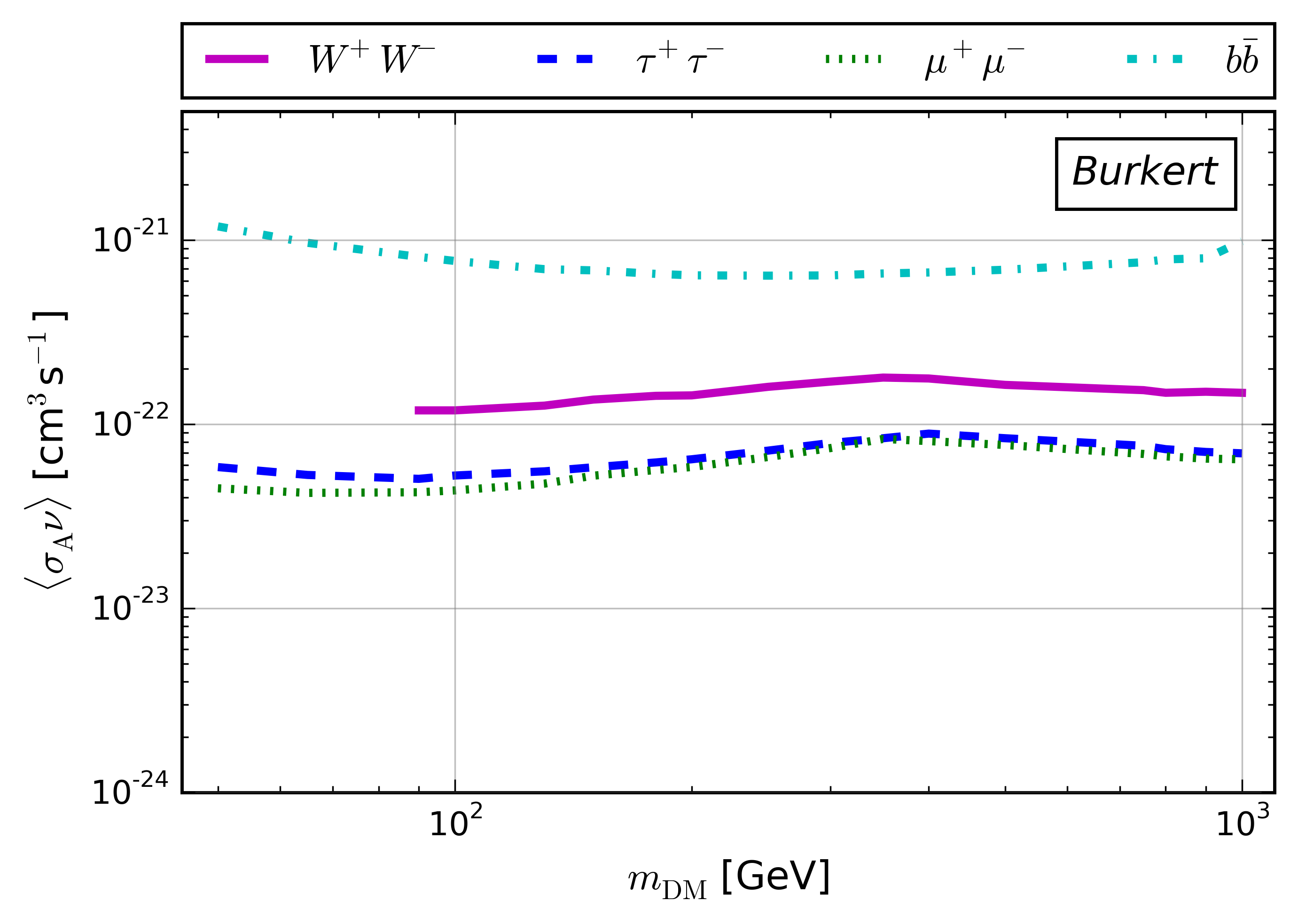}
    \end{minipage}
    \caption{Combined 90$\%$ CL limits on the thermally-averaged dark matter annihilation cross section as a function of the dark matter mass for the NFW (top) and Burkert (bottom) halo profiles. All annihilation channels considered in this analysis are presented ($b\bar{b}$, $\tau^+\tau^-$, $\mu^+\mu^-$, $W^+W^-$).}
    \label{fig:AllChannels_Limits}
\end{figure}

This joint analysis is conducted with data collected by the ANTARES and IceCube neutrino telescopes during a period of 9 and 3 years, respectively. By combining the data samples at the likelihood level, we find no significant excess of neutrinos in the direction of the Galactic Centre. We present limits on the thermally-averaged dark matter self-annihilation cross section $\langle \sigma_A \upsilon \rangle$. The values obtained for all dark matter masses and annihilation channels can be found in Tables~\ref{tab:NFW_results} and~\ref{tab:Burkert_results} for the NFW and Burkert profiles, respectively, with parameters from Table~\ref{tab:ModelParam}. 
The $90\%$ CL combined limits are presented in Figure~\ref{fig:AllChannels_Limits} for all self-annihilation channels considered, assuming both the NFW (top) and Burkert (bottom) halo profiles. 
The dissimilar behaviour of these limits for the NFW and Burkert profiles originates from the two event reconstructions used for ANTARES. The transition point 

%Limits all channels

\noindent from QFit to $\lambda$Fit depends on the balance between the number of reconstructed events and the quality of the reconstruction. The better angular resolution provided by $\lambda$Fit is more beneficial when considering a ``cuspy'' dark matter halo profile such as the NFW profile. Therefore, the transition between the two reconstruction happens at a lower dark matter mass for the NFW profile.

% Compare to other experiments
\begin{center}
\begin{figure}
\includegraphics[width=0.99\linewidth]{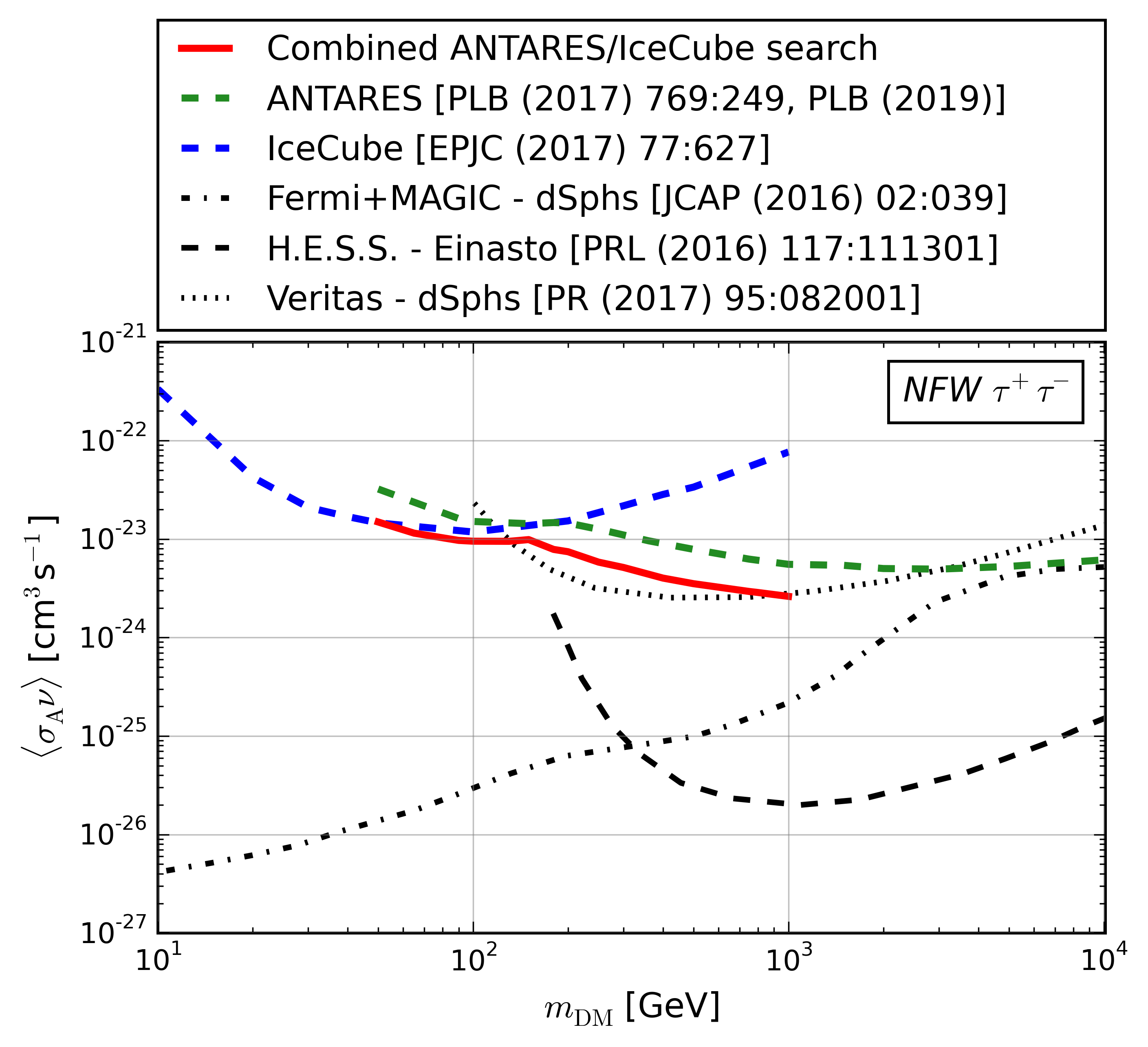}
\caption{90\% CL upper limit on the thermally-averaged dark matter annihilation cross section $\langle\sigma_A\upsilon\rangle$ obtained for the combined analysis as a function of the dark matter mass $m_{\mathrm{DM}}$ assuming the NFW halo profile for the $\tau^+\tau^-$ annihilation channel. The limits from IceCube~\cite{IC86_GCWIMP}, ANTARES~\cite{ANTARES_GCWIMP}, VERITAS~\cite{Veritas}, Fermi+MAGIC~\cite{Fermi-MAGIC} and H.E.S.S.~\cite{HESS} are also shown.}
\label{fig:limits_ comparison}
\end{figure}
\end{center}

\begin{center}
\begin{figure}
\includegraphics[width=0.99\linewidth]{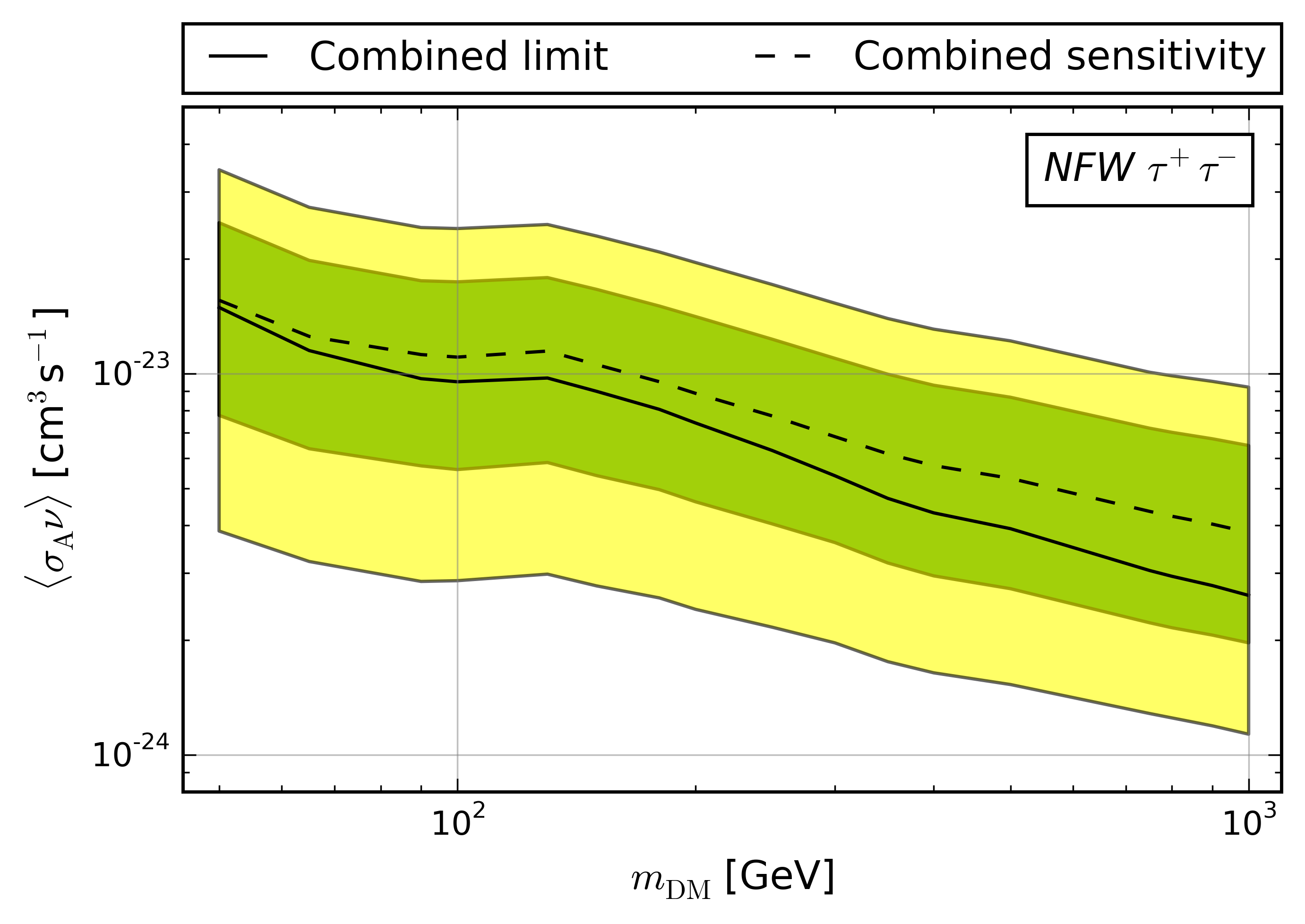}
\caption{Comparison of the 90$\%$ CL combined limit (solid line) and sensitivity (dashed line) for the NFW halo profile and the $\tau^+\tau^-$ annihilation channel, along with the expected $1\sigma$ (green) and $2\sigma$ (yellow) bands around the expected median sensitivity.}
\label{fig:brazilian_plot}
\end{figure}
\end{center}

In Figure~\ref{fig:limits_ comparison}, we present the combined limit obtained for the $\tau^+\tau^-$ channel and the NFW profile alongside the previous ANTARES and IceCube limits. The present analysis uses the datasets developed for these individual searches. When compared to the IceCube and ANTARES stand-alone limits, the combined limit is better by up to a factor 2 in the dark matter mass range considered, \textit{i.e.} between 50 and 1000\,GeV. 
An enhancement of the limit can also be seen for the other dark matter annihilation channel and halo profile combinations presented in Figure~\ref{fig:AllChannels_Limits}, with an exception for the $b\bar{b}$ channel when considering the Burkert profile. For this particular case, the combined limit is dominated by IceCube, which has a better signal acceptance than ANTARES for the entire mass range due to the very soft spectrum.
In addition to the improvement due to the combination of the two datasets, a difference between the ANTARES limit and the combined limit is also noticeable for dark matter masses close to 1 TeV, where the contribution from IceCube is expected to be negligible. This divergence results mainly from the way under-fluctuations are treated by this analysis and the previous ANTARES search. When obtaining limits with lower values than sensitivities, sensitivities were labelled as limits for the previous ANTARES analysis while limits remain unchanged for our combined search. Furthermore, the ANTARES analysis used the Neyman approach~\cite{Neyman} with slightly different PDFs for the $\lambda$Fit reconstruction. The importance of these changes can be seen in Figure~\ref{fig:brazilian_plot}, where the limit for dark matter annihilation into $\tau^+\tau^-$ for the NFW profile is shown alongside the sensitivity. 
These results are also compared with current limits obtained with \gammaray{} telescopes from searches of photons produced in the self-annihilation of dark matter into $\tau^+\tau^-$ (see Figure~\ref{fig:limits_ comparison}). Gamma-ray limits are still several order of magnitude better for this particular channel although it needs to be noted that the VERITAS~\cite{Veritas} and the combined Fermi+MAGIC limits~\cite{Fermi-MAGIC} were obtained from the study of dwarf spheroidal galaxies (dSphs), while the other limits presented are for the Galactic Centre.
Note as well that the H.E.S.S. limit was obtained assuming the Einasto halo profile~\cite{HESS}.
Although both the NFW and Einasto halo profiles assume a high dark matter density at the centre of the galaxy, the difference between the profiles is non-negligible in the central region. Moreover, there is considerable freedom in the choice of halo parameters, and these choices are not made consistently between experiments. The halo parameters used in this work are conservative with respect to more optimistic values made in other analyses, and this freedom is responsible for some of the difference between the limits set by IceCube and the more stringent limits reported in~\cite{HESS}.

%Result table NFW
\begin{center}
\begin{table}
    \begin{tabular}{c c c c c}
    \hline
    $m_{\mathrm{DM}}$ [GeV] & \multicolumn{4}{c}{$\langle\sigma_A\upsilon\rangle$ [$10^{-24}$ cm$^3$s$^{-1}$]} \\
    \hline
     & $b\bar{b}$ & $\tau^+\tau^-$ & $\mu^+\mu^-$ & $W^+W^-$   \\
    \hline
    50 & 424 & 14.9 & 12.3 & ---  \\
    65 & 315 & 11.5 & 9.8 & ---  \\
    90 & 236 & 9.7 & 8.3 & 23.5  \\
    100 & 217 & 9.5 & 8.2 & 21.7   \\
    130 & 177 & 9.5 & 8.2 & 21.5   \\
    150 & 162 & 9.9 & 8.7 & 22.9   \\
    180 & 157 & 7.9 & 6.9 & 15.0   \\
    200 & 144 & 7.4 & 6.3 & 13.6   \\
    250 & 136 & 5.8 & 5.2 & 10.6 \\
    300 & 132 & 5.2 & 4.5 & 9.5 \\
    350 & 130 & 4.5 & 4.2 & 8.1   \\
    400 & 131 & 4.0 & 3.6 & 7.8   \\
    500 & 107 & 3.5 & 3.3 & 6.9   \\
    750 & 72.9 & 2.9 & 2.7 & 5.8   \\
    800 & 69.7 & 2.9 & 2.6 & 5.9   \\
    900 & 66.0 & 2.7 & 2.4 & 5.7   \\
    1000 & 66.1 & 2.6 & 2.4 & 5.5  \\
    \hline
    \end{tabular}
\caption{90$\%$ CL upper limits on the thermally-averaged self-annihilation cross section for the NFW profile.}
\label{tab:NFW_results}
\end{table}
\end{center}

%Result table Burkert
\begin{center}
\begin{table}
    \begin{tabular}{c c c c c}
    \hline
    $m_{\mathrm{DM}}$ [GeV] & \multicolumn{4}{c}{$\langle\sigma_A\upsilon\rangle$ [$10^{-23}$ cm$^3$s$^{-1}$]} \\
    \hline
     & $b\bar{b}$ & $\tau^+\tau^-$ & $\mu^+\mu^-$ & $W^+W^-$   \\
    \hline
    50 & 118 & 5.9 & 4.5 & ---   \\
    65 & 96.8 & 5.3 & 4.2 &  ---  \\
    90 & 81.2 & 5.1 & 4.3 & 11.9 \\
    100 & 77.1 & 5.3 & 4.4 &  11.9  \\
    130 & 69.6 & 5.6 & 4.8 &  12.6  \\
    150 & 68.6 & 5.9 & 5.3 &  13.6  \\
    180 & 65.7 & 6.2 & 5.7 &  14.3  \\
    200 & 64.5 & 6.5 & 5.9 &  14.4  \\
    250 & 64.2 & 7.2 & 6.7 &   15.9 \\
    300 & 64.4 & 7.9 & 7.4 &  17.1  \\
    350 & 65.9 & 8.4 & 8.3 &  17.9  \\
    400 & 66.8 & 8.9 & 8.1 &  17.8  \\
    500 & 69.1 & 8.4 & 7.7 &  16.4  \\
    750 & 75.9 & 7.7 & 6.9 &  15.3  \\
    800 & 78.7 & 7.3 & 6.7 &  14.8  \\
    900 & 79.8 & 7.1 & 6.5 &  15.0  \\
    1000 & 98.7 & 7.0 & 6.5 & 14.8 \\
    \hline
    \end{tabular}
\caption{90$\%$ CL upper limits on the thermally-averaged self-annihilation cross section for the Burkert profile.}
\label{tab:Burkert_results}
\end{table}
\end{center}

\section*{Acknowledgements}
%%ANTARES%%
The authors of ANTARES collaboration acknowledge the financial support of the funding agencies:
% France:
Centre National de la Recherche Scientifique (CNRS), Commissariat \`a
l'\'ener\-gie atomique et aux \'energies alternatives (CEA),
Commission Europ\'eenne (FEDER fund and Marie Curie Program),
Institut Universitaire de France (IUF), IdEx program and UnivEarthS
Labex program at Sorbonne Paris Cit\'e (ANR-10-LABX-0023 and
ANR-11-IDEX-0005-02), Labex OCEVU (ANR-11-LABX-0060) and the
A*MIDEX project (ANR-11-IDEX-0001-02),
R\'egion \^Ile-de-France (DIM-ACAV), R\'egion
Alsace (contrat CPER), R\'egion Provence-Alpes-C\^ote d'Azur,
D\'e\-par\-tement du Var and Ville de La
Seyne-sur-Mer, France;
% Germany: 
Bundesministerium f\"ur Bildung und Forschung
(BMBF), Germany; 
% Italy
Istituto Nazionale di Fisica Nucleare (INFN), Italy;
% Netherlands
Nederlandse organisatie voor Wetenschappelijk Onderzoek (NWO), the Netherlands;
% Russia
Council of the President of the Russian Federation for young
scientists and leading scientific schools supporting grants, Russia;
% Romania
Executive Unit for Financing Higher Education, Research, Development and Innovation (UEFISCDI), Romania;
% Spain
Ministerio de Ciencia, Innovaci\'{o}n, Investigaci\'{o}n y Universidades (MCIU): Programa Estatal de Generaci\'{o}n de Conocimiento (refs. PGC2018-096663-B-C41, -A-C42, -B-C43, -B-C44) (MCIU/FEDER), Severo Ochoa Centre of Excellence and MultiDark Consolider (MCIU), Junta de Andaluc\'{i}a (ref. SOMM17/6104/UGR), 
Generalitat Valenciana: Grisol\'{i}a (ref. GRISOLIA/2018/119), Spain;
% Marocco
Ministry of Higher Education, Scientific Research and Professional Training, Morocco.
% A.O.B.:
We also acknowledge the technical support of Ifremer, AIM and Foselev Marine
for the sea operation and the CC-IN2P3 for the computing facilities.

%%IceCube%%
The authors of IceCube collaboration gratefully acknowledge the support from the following agencies and institutions:
%USA
USA {\textendash} U.S. National Science Foundation-Office of Polar Programs,
U.S. National Science Foundation-Physics Division,
Wisconsin Alumni Research Foundation,
Center for High Throughput Computing (CHTC) at the University of Wisconsin-Madison,
Open Science Grid (OSG),
Extreme Science and Engineering Discovery Environment (XSEDE),
U.S. Department of Energy-National Energy Research Scientific Computing Center,
Particle astrophysics research computing center at the University of Maryland,
Institute for Cyber-Enabled Research at Michigan State University,
and Astroparticle physics computational facility at Marquette University;
%Belgium
Belgium {\textendash} Funds for Scientific Research (FRS-FNRS and FWO),
FWO Odysseus and Big Science programmes,
and Belgian Federal Science Policy Office (Belspo);
%Germany
Germany {\textendash} Bundesministerium f{\"u}r Bildung und Forschung (BMBF),
Deutsche Forschungsgemeinschaft (DFG),
Helmholtz Alliance for Astroparticle Physics (HAP),
Initiative and Networking Fund of the Helmholtz Association,
Deutsches Elektronen Synchrotron (DESY),
and High Performance Computing cluster of the RWTH Aachen;
%Sweden
Sweden {\textendash} Swedish Research Council,
Swedish Polar Research Secretariat,
Swedish National Infrastructure for Computing (SNIC),
and Knut and Alice Wallenberg Foundation;
%Australia
Australia {\textendash} Australian Research Council;
%Canada
Canada {\textendash} Natural Sciences and Engineering Research Council of Canada,
Calcul Qu{\'e}bec, Compute Ontario, Canada Foundation for Innovation, WestGrid, and Compute Canada;
%Denmark
Denmark {\textendash} Villum Fonden, Danish National Research Foundation (DNRF), Carlsberg Foundation;
%New Zealand
New Zealand {\textendash} Marsden Fund;
%Japan
Japan {\textendash} Japan Society for Promotion of Science (JSPS)
and Institute for Global Prominent Research (IGPR) of Chiba University;
%Korea
Korea {\textendash} National Research Foundation of Korea (NRF);
%Switzerland
Switzerland {\textendash} Swiss National Science Foundation (SNSF);
%UK
United Kingdom {\textendash} Department of Physics, University of Oxford.
The IceCube collaboration acknowledges the significant contributions to this manuscript from Sebastian Baur, Nad\`{e}ge Iovine and Sara Rebecca Gozzini.

\bibliography{references}

\end{document}